\documentclass[graybox]{svmult}

\usepackage{type1cm}        
\usepackage{makeidx}         
\usepackage{graphicx}        
\usepackage{multicol}        
\usepackage[bottom]{footmisc}

\usepackage{newtxtext}       %
\usepackage[varvw]{newtxmath}       

\makeindex             

\begin{document}

\title*{Increasing quantum speed limit of relativistic electron via 
non-uniform magnetic field
}
\titlerunning{Increasing quantum speed limit via non-uniform magnetic field}
\author{Srishty Aggarwal\orcidID{0000-0001-5809-1994}
\\Banibrata Mukhopadhyay\orcidID{0000-0002-3020-9513}
\\Subhashish Banerjee\orcidID{0000-0002-7739-4680} 
\\Arindam Ghosh\orcidID{0000-0001-5188-4617} and
\\Gianluca Gregori\orcidID{0000-0002-4153-0628}}
\authorrunning{Aggarwal, Mukhopadhyay, Banerjee, Ghosh \& Gregori}
\institute{Srishty Aggarwal \at Department of Physics, Indian Institute of Science, Bengaluru, India 560012  \email{srishtya@iisc.ac.in}
\and Banibrata Mukhopadhyay  \at Department of Physics, Indian Institute of Science, Bengaluru, India 560012 \email{bm@iisc.ac.in}
\and Subhashish  Banerjee \at Department of Physics, Indian Institute of Technology Jodhpur, Jodhpur, India 342030 \email{subhashish@iitj.ac.in}
\and Arindam Ghosh  \at Department of Physics, Indian Institute of Science, Bengaluru, India 560012 \email{arindam@iisc.ac.in}
\and Gianluca Gregori  \at Department of Physics, University of Oxford, Oxford OX1 3PU, UK \email{gianluca.gregori@physics.ox.ac.uk}
}
%
%
\maketitle
\vspace{1 cm}
\textit{To be published in 
Astrophysics and Space Science Proceedings, titled "The Relativistic Universe: From Classical to Quantum, Proceedings of the International Symposium on Recent Developments in Relativistic Astrophysics", Gangtok, December 11-13, 2023: to felicitate Prof. Banibrata Mukhopadhyay on his 50th Birth Anniversary", Editors: S Ghosh \& A R Rao, Springer Nature.}

\vspace{5 cm}

\abstract{
Quantum speed limit (QSL) defines the theoretical upper bound on how fast a quantum system can evolve between states. It imposes a fundamental constraint on the rate of quantum information processing. For a relativistic spin-up electron in a uniform magnetic field, QSL increased with the magnetic field strength till around $10^{15}$ Gauss, before saturating at a saturated QSL (SQSL) of $0.2407c$, where ‘$c$’ is the speed of light. We show that by using variable magnetic fields, it is possible to surpass this limit, achieving SQSL upto $0.4-0.6c$. To attain this quantum phenomenon, we solve the evolution equation of relativistic electron in spatially varying magnetic fields and find that the energies of various electron states become non-degenerate as opposed to the constant magnetic field case. This redistribution of energy is the key ingredient to accomplish higher QSL and, thus, a high information processing speed. We further explore how QSL can serve as a bridge between relativistic and non-relativistic quantum dynamics, providing insights via the Bremermann-Bekenstein bound,  a quantity which constrains the maximal rate of information production. We also propose a practical experimental setup to realize these advancements. These results hold immense potential for propelling fields of quantum computation, thermodynamics and metrology.}

\section{Introduction}
\label{sec:1}
The quantum speed limit (QSL) imposes an elemental constraint on the rate at which a quantum system can evolve. This limit, rooted in the Heisenberg uncertainty principle, is pivotal in quantum algorithms and information processing speed \cite{Deffner2017}. 

The QSL concept has illuminated diverse facets of quantum science, including information processing, open system dynamics, quantum control, and quantum thermodynamics, where precise manipulation and measurement of quantum systems are essential. Further, QSL time can be related to the energy cost per bit of information via Bremermann-Bekenstein (BB) bound \cite{Bremermann,Bekenstein1990}.

Magnetic fields are crucial for controlling quantum systems. Scientists have used uniform magnetic fields to manipulate chains of molecular qubits \cite{Santini}. The QSL of relativistic electron reached to $0.2407c$,  when subjected to uniform magnetic field ($c$ is the speed of light) \cite{villamizar}. Can non-uniform magnetic fields be used to increase QSL, thereby accelerating quantum information processing?

We, therefore, investigate the QSL for a relativistic electron in a superposition of its ground and first excited states under spatially varying magnetic field. Further, we identify the critical magnetic field 
representing the boundary between non-relativistic and relativistic regimes by using the BB bound. In the end, we propose a feasible experimental setup to generate such magnetic field configurations.

\section{Energy levels of relativistic electron in non-uniform magnetic field}
\label{sec:2}

We utilize a strictly spatially varying magnetic field profile, as detailed in \cite{SciSris}, which is consistent with Maxwell's electromagnetic laws and given by
\begin{equation}
\textbf{B} = B_{0}\rho^n \hat{z},
\label{eqB}
\end{equation}
in cylindrical coordinates $(\rho,\phi,z)$. The magnetic field is uniform if $n=0$, and is strictly spatially increasing or decreasing depending on whether $n>0$ or $n<0$ respectively. However, the solution of Eq. (\ref{eq3}) is converging only till $n>-1$ \cite{SciSris}.

We choose the vector potential \textbf{A} as
\begin{equation}
 \textbf{A} = B_{0}\frac{\rho^{n+1}}{n+2} \hat{\phi} = A\hat{\phi}.
\end{equation}
To investigate the dynamics of a relativistic electron within a non-uniform magnetic field, we solve the Dirac equation, given by
\begin{equation}
i\hbar\frac{\partial\Psi}{\partial t} = \left[ c\boldsymbol{\alpha}\cdot\left(-i\hbar\textbf{$\nabla$}-\frac{q\textbf{A}}{c}\right) + \beta m_{e}c^2\right]\Psi,
\label{eqn1}
\end{equation}
where $m_{e}$ and $q~(-e)$ are electron's mass and charge respectively, $\hbar$ is the reduced Planck constant, $\textbf{A}$ 
is the vector potential and $\boldsymbol{\alpha}$ and $\beta$ are Dirac matrices. For stationary states, the wavefunction can be written as
\begin{equation}
	\Psi = e^{-i\frac{Et}{\hbar}}\begin{bmatrix}
  \chi \\
 \phi \\
 \end{bmatrix}
\label{matrix},
 \end{equation} 
$\phi$ and $\chi$ are 2-component spinors. The antiparticle wavefunction $\phi=-\chi$ \cite{1998rqm..book.....S}. 

Defining $\boldsymbol{\pi}=-i\hbar\textbf{$\nabla$}-q\textbf{A}/c$ and using $(\boldsymbol\sigma\cdot\boldsymbol\pi)(\boldsymbol\sigma\cdot\boldsymbol\pi) = \pi ^2 - q\hbar\boldsymbol\sigma\cdot \boldsymbol{B}/c$, the  decoupled equation obtained for $\chi$ is
\begin{equation}
(E^2-m_{e}^2c^4)\chi = \left[c^2\left(\pi ^2 - \frac{q\hbar}{c}\boldsymbol\sigma\cdot \textbf{B}\right)\right]\chi. 
\label{eq3}
\end{equation}
Let
\begin{equation}
 \chi = e^{i\left(m\phi+\frac{p_{z}}{\hbar}z\right)}R(\rho),
 \label{eq4}
 \end{equation}
where `$m\hbar$' is the angular momentum, $p_{z}$ is the eigenvalue of momentum in the $z-$direction and $R(\rho)$ is a two-component matrix. $p_z=0$ as the electron is confined to a plane perpendicular to the $z$-direction. Substituting $\chi$ from Eq. (\ref{eq4}) into Eq. (\ref{eq3}), and dividing it by $m_{e}^2 c^2$, Eq. (\ref{eq3}) becomes

\begin{align}
\alpha\tilde{R}_{\pm}  &= -\lambda_{e}^2\left[\frac{\partial ^2}{\partial \rho^2}+\frac{1}{\rho}\frac{\partial}{\partial\rho}-\frac{m^2}{\rho^2}\right]\tilde{R}_{\pm}
\nonumber \\
&\quad  + \left[\left(\frac{kB_0\rho^{n+1}}{n+2}\right)^2 + k\lambda_{e}\left(-\frac{2m}{n+2}\pm 1\right)B_0\rho^n\right]\tilde{R}_{\pm},
\label{eq10}
\end{align}
$\alpha$ is the eigenvalue of the problem and is given by $\epsilon^2-1$, $\epsilon = E/m_{e}c^2$ (dimensionless energy),
$\lambda_{e} = \hbar/m_{e}c$ (Compton wavelength of electrons),
$k=e/m_{e}c^2$ \cite{SciSris}.
Hence, energy of level $\nu$ becomes
\begin{equation}
    E_{\nu} = m_e c^2 \sqrt{1+\alpha_{\nu}}.
    \label{EalphaEq}
\end{equation}
In absence of Zeeman splitting and with $m=0$, Eq. (\ref{eq10}) becomes
\begin{equation}
\alpha'\tilde{R}  = -\lambda_{e}^2\left[\frac{\partial ^2}{\partial \rho^2}+\frac{1}{\rho}\frac{\partial}{\partial\rho}-\right]\tilde{R}+ \left(\frac{kB_0\rho^{n+1}}{n+2}\right)^2 \tilde{R},
\label{eq11}
\end{equation}
where $\alpha'$ is the energy level excluding Zeeman effect. 

\begin{figure}

\includegraphics[scale=.35]{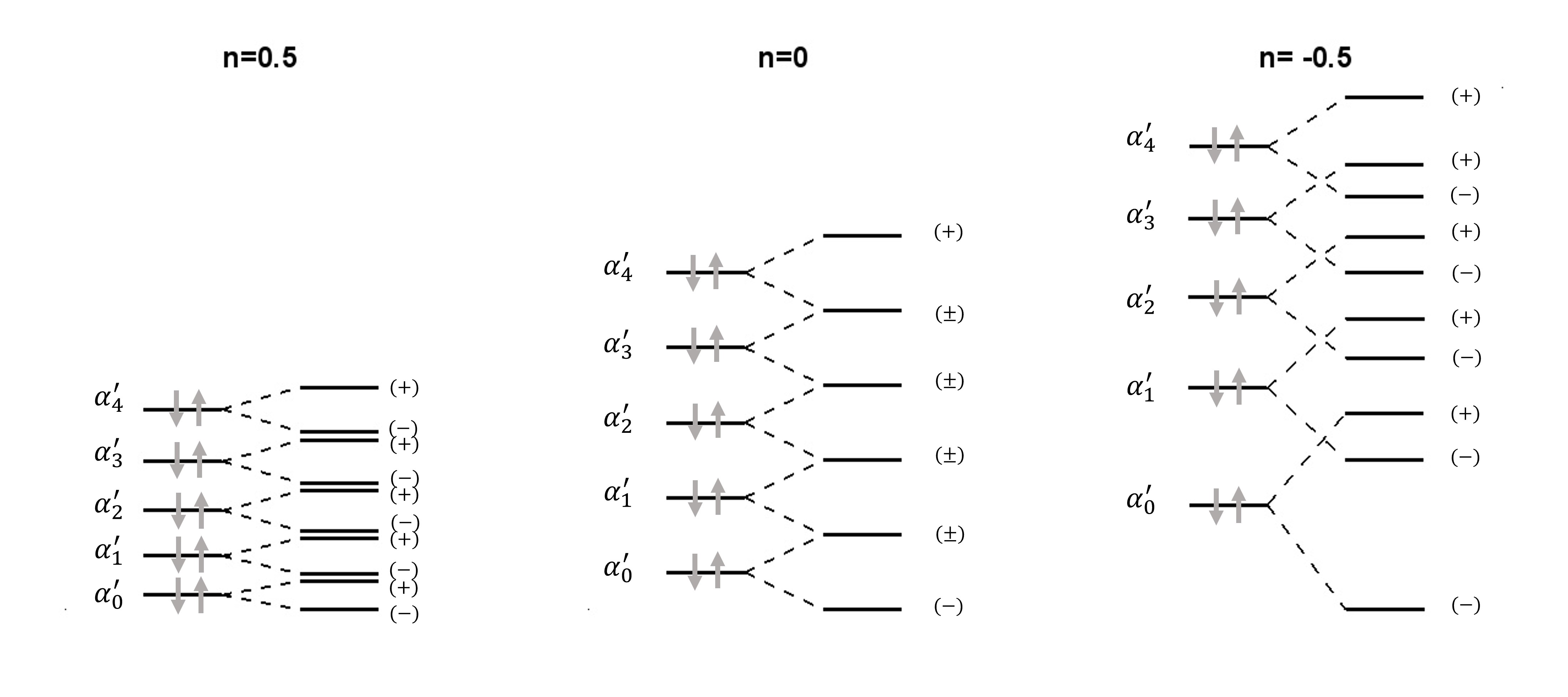}
%
%
\caption{Schematic representation for comparison of first five eigenlevels before splitting ($\alpha'_{\nu}$) and after splitting into $-\sigma.B$ and $+\sigma.B$ taking $m=0$ for $n=0.5,\: 0$ and $n=-0.5$.}
\label{fig:1}       
\end{figure}

Solving Eq. (\ref{eq10}) using Runga Kutta method \cite{SciSris}, we can obtain the eigenspectrum for different $n$. The mathematical relation for eigenvalues with $n$ has been shown in {\cite{MGSris}. The eigenspectrum, corresponding to $\alpha'_{\nu}$, and the split states $\alpha_{\nu}$ for first five levels are given in Fig. \ref{fig:1}. As depicted, with increasing eigenstates, eigenvalue difference for $\alpha'_{\nu}$ between the two consecutive states stays the same for $n=0$, increases for $n>0$ and decreases for $n<0$. The main reason for this difference is that in first case (growing magnetic field), electron observes a weak magnetic field close to the center, thereby, has a lesser gap in lower levels. As the field strength increases, the discretion in the energy levels increases and, therefore, the gap between the energy levels increases in case of positive $n$. Similar reasoning holds for negative $n$ leading to an initial large gap in the eigenlevels and its subsequent decrease.

Moreover, the spin degeneracy inherent in a uniform magnetic field is broken in the case of a non-uniform field. One of the important differences between the radially growing and decaying magnetic fields is the corresponding alignment of spin-down and spin-up energy levels. In the case of a growing magnetic field (positive $n$), the lower energy level of the spin-up electron lies below the subsequent energy level of the spin-down electron. On the other hand, for decaying magnetic field (negative $n$), the former is positioned above the latter, as is illustrated for $n=0.5$ and $-0.5$ in Fig. \ref{fig:1}. These different alignments can be used to generate different closed systems of interests. For example, using negative $n$, one can generate a system of spin-down electrons confined to the gound and first excited states. Similarly, one can build a two-level system with only spin transitions in case of positive $n$.

\section{QSL in the non-uniform magnetic field}
\label{sec:4}

In general, for the evolution from one state to the other, the QSL of an electron ($\tilde{v}$) can be evaluated as the ratio of the its radial displacement ($\rho_{disp}$) to the minimum time of evolution ($\tau_{QSL}$), given by

\begin{equation}
\tilde{v} = \frac{\rho_{disp}}{\tau_{QSL}}.
\label{eq vel}
\end{equation}
The initial ($t=0$) and final ($t=\tau_{QSL}$) states of electron, assuming it to be in the superposition of the $\nu_{th}$ and $(\nu +1)_{th}$ states at all times, are
\begin{equation}
\Psi(r,0) = \frac{1}{\sqrt{2}}\left[\psi_{\nu}(r)+\psi_{\nu+1}(r)\right],
\end{equation}
and
\begin{equation}
\Psi(r,\tau_{QSL}) = \frac{1}{\sqrt{2}}\left[\psi_{\nu}(r)e^{\frac{iE_{\nu}\tau_{QSL}}{\hbar}}+\psi_{\nu+1}(r)e^{\frac{iE_{\nu+1}\tau_{QSL}}{\hbar}}\right].
\end{equation}
$\tau_{QSL}$ between two orthogonal states is
\begin{equation}
    \tau_{QSL} = max \left\{\frac{\pi \hbar}{2\Delta H},\frac{\pi \hbar}{2\langle H\rangle} \right\},
\end{equation}
where the first and second terms are the Mandelstam-Tamm (MT) \cite{mandelstam1945energy} and Margolous-Levitian (ML) \cite{MargolousLevitin} bounds respectively. As all the energy levels are positive, $\Delta H$ is smaller than $\langle H \rangle$. Therefore, $\tau_{QSL}$ is given by MT bound. Hence,
\begin{equation}
\tau_{QSL}=\frac{\pi \hbar}{2\Delta H},
\label{eqn4}
\end{equation}
with
\begin{equation}
\Delta H =\frac{ E_{\nu +1}-E_{\nu}}{2}.
\end{equation}
The mean radial position at a time $t$ is represented by
\begin{equation}
\langle\rho\rangle = \frac{1}{2}\left[\langle\nu|\rho|\nu\rangle + \langle\nu+1|\rho|\nu+1\rangle + 2\langle\nu|\rho|\nu+1\rangle cos({\cal{E}}t)\right],
\end{equation}
where 
\begin{equation}
\nonumber
{\cal{E}} = \frac{E_{\nu +1}-E_{\nu}}{\hbar}.
\end{equation}
Therefore,
\begin{align}
\rho_{disp} &= |\langle\rho\rangle_{\tau_{QSL}}-\langle\rho\rangle_0|\nonumber \\ 
&=2\left\vert\int^\infty_0 \rho D_S(\rho)d\rho\: \right\vert,
\label{eqn5}
\end{align}
where 
\begin{equation}
D_s(\rho)=\psi^{\dagger}_{\nu}\:\rho\:\psi_{\nu+1}
\label{eqn6}
\end{equation}
Thus, QSL can be evaluated as the ratio of Eq. (\ref{eqn5}) to Eq. (\ref{eqn4}) \cite{NJPSris}.

\begin{figure}
\includegraphics[scale=.35]{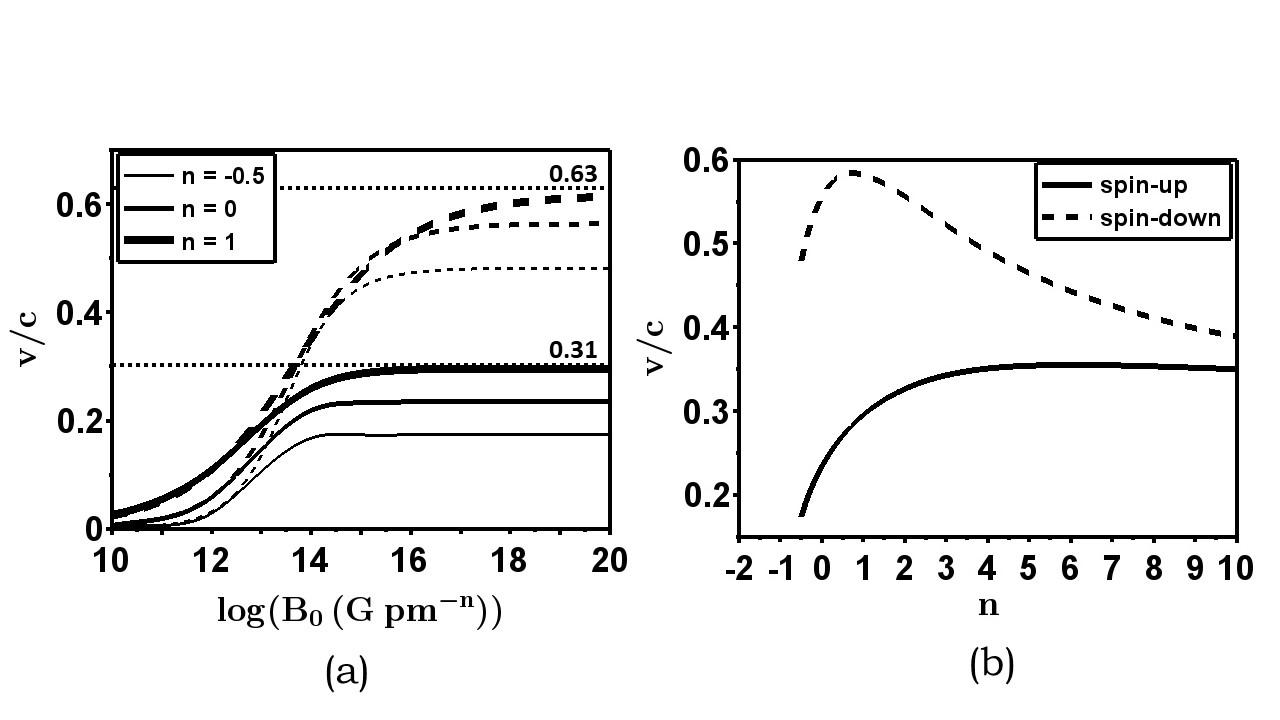}

\caption{(a) Variation of QSL with magnetic field strength ($B_0$) for increasing ($n=1$), uniform ($n=0$) and decreasing ($n= -0.5$) magnetic field profile for spin-up (solid lines) and spin-down (broken lines) electron. The line thickness is in order of increasing $n$. Here, for $n=1$, SQSLs are highlighted by horizontal line. (b) Variation of QSL with $n$ at $B_0 = 10^{17}\:G\:pm^{-n}$.}
\label{fig:2}       
\end{figure}

The variation of QSL with the magnetic field strength is shown in Fig. \ref{fig:2} (a) for uniform ($n=0$), decreasing ($n=-0.5$) and increasing ($n=1$) fields for spin-up and spin-down electrons. QSL increases with the increase in $B_0$ and then tends to saturated QSL (SQSL). The SQSL is higher for $n=1$ as compared to $n=0$ and $-0.5$. To determine the change of SQSL with $n$, the variation of QSL with $n$ is depicted in Fig. \ref{fig:2} (b) at very high magnetic field strength of $B_0 = 10^{17}\: G\:pm^{-n}$, such that QSL tends to SQSL. The SQSL tends to increase with $n$ for spin-up electron, while for spin-down electron, it increases up to $n\thicksim2$, and then decreases with $n$. Thus growing magnetic field type variation leads to higher QSL as compared to uniform magnetic field.

\section{Bremermann-Bekenstein bound: finding critical magnetic field}
\label{sec:5}

The relativistic treatment becomes important only when the energy associated with magnetic field exceeds the rest mass energy of the particle. The magnetic field threshold beyond which the relativistic dynamics dominates is known as the critical magnetic field. For electrons, the critical magnetic field is given by $m_e^2c^3/\hbar e$ Gauss in case of a uniform magnetic field and achieved when the gyromagnetic radius approximates its Compton wavelength. However, such an analysis is much more involved for a non-uniform magnetic field. Interestingly, we can find the critical magnetic field for the latter case using a unique tool of quantum information: the BB bound. 

BB bound is a fundamental theoretical limit on the fastest possible speed of information processing or storage. It provides a link between the energy cost $\langle H \rangle$ for a bit of information $I$ and the QSL time, and is given by \cite{deffner}
\begin{equation}
\frac{\langle H \rangle}{I} > \frac{\hbar \ln 2}{ \pi \tau_{QSL}}
\label{BB bound}.   
\end{equation}
 This bound arises from the intersection of quantum mechanics, general relativity, and information theory, offering deep insights into the fundamental constraints of physical systems \cite{Bremermann}. It has profound implications in quantum information, cosmology and black hole physics. It also leads to the development of concepts such as black hole entropy and the holographic principle.

\begin{figure}[t]
\sidecaption[t]
\includegraphics[scale=.45]{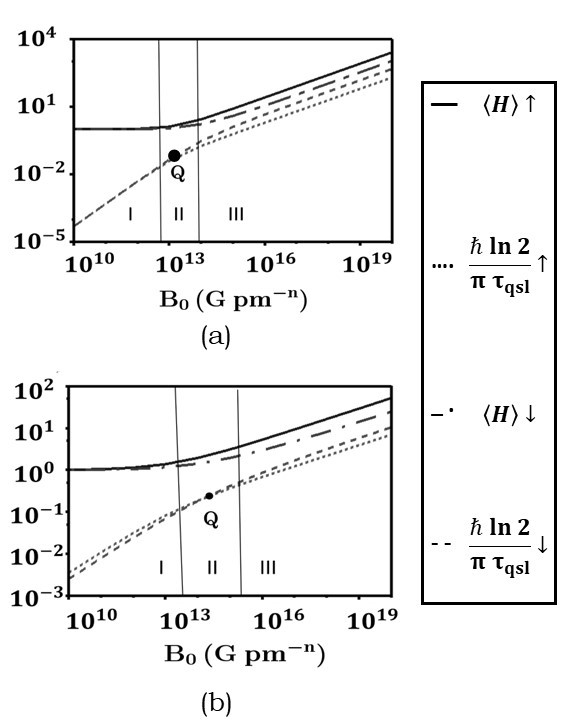}
%
%
\caption{Variation of $\langle H \rangle$ (LHS) and $\hbar \:ln \:2 / \pi \tau_{QSL}$ (RHS) of Eq. (\ref{BB bound}) with $B_0$ for spin-up ($\uparrow$) and spin-down ($\downarrow$) electrons as an illustration of the difference between non-relativistic and relativistic regimes for (a) $n=0$ and (b) $n=2$. I, II and III regions represent `non-relativistic', `transition' and `relativistic' regimes, respectively. \textit{Q} represents the critical magnetic field. While in case of uniform magnetic field in (a), $(\hbar ln2)/(\pi\tau_{QSL})$ is same for both the spins, in non-relativistic regime, it is higher for spin-down electron in the relativistic regime. Further, it is smaller for the spin-down electron in non-relativistic regime in the case of non-uniform magnetic field (b), while larger in the relativistic one.}
\label{fig:Bstability}       
\end{figure}

 Fig. \ref{fig:Bstability} shows the variation of LHS (taking $I=1$) and RHS of Eq. (\ref{BB bound}) with $B_0$ for two cases: (a) $n=0$ and (b) $n=2$. The two sub-figures are split into three regions. In region I, characterized by low $B_0$ and non-relativistic dynamics,  $\langle H\rangle$ and $(\hbar ln2)/(\pi\tau_{QSL})$ exhibit a clear separation. As we transition to region II, with a stronger magnetic field, this separation diminishes, indicating a shift towards relativistic behavior. In region III, dominated by a strong magnetic field, the two quantities become nearly parallel, signifying a fully relativistic regime. Thus, the difference in the regimes of non-relativistic and relativistic is indeed evident. 

The point of separation of RHS for spin-down and spin-up electrons, highlighted as \textbf{Q} ($4.414\times 10^{13}\:G$) in Fig. \ref{fig:Bstability} (a), depicts the already known critical magnetic field for uniform case ($n=0$). In Fig. \ref{fig:Bstability} (b), \textbf{Q} ($B_0 = 1.35\times10^{14}\:G\:pm^{-n}$)  is the point of intersection for spin-down and spin-up electrons. Using a similar comparison for non-uniform magnetic field, \textbf{Q} may indicate the critical magnetic field for $n=2$. Hence, BB bound furnishes an easy technique for estimating the critical magnetic field.

\section{Experimental design for non-uniform magnetic field}
\label{sec:3}

Generating non-uniform magnetic fields requires customised approaches based on the specific quantum system under investigation. To attain varying magnetic field, a solenoid with its ferrite core equipped with curved pole pieces can be taken as shown in Fig. \ref{fig:experiment}. Utilizing high-permeability materials like ferrite or soft iron for the core enhances field strength. The concavity (convexity) of the pole pieces will lead to smaller (larger) magnetic field strength at the center, and larger (smaller) at the periphery. Thus, the curvature can be adjusted to attain the desired $n$. 

\begin{figure}
\includegraphics[scale=.4]{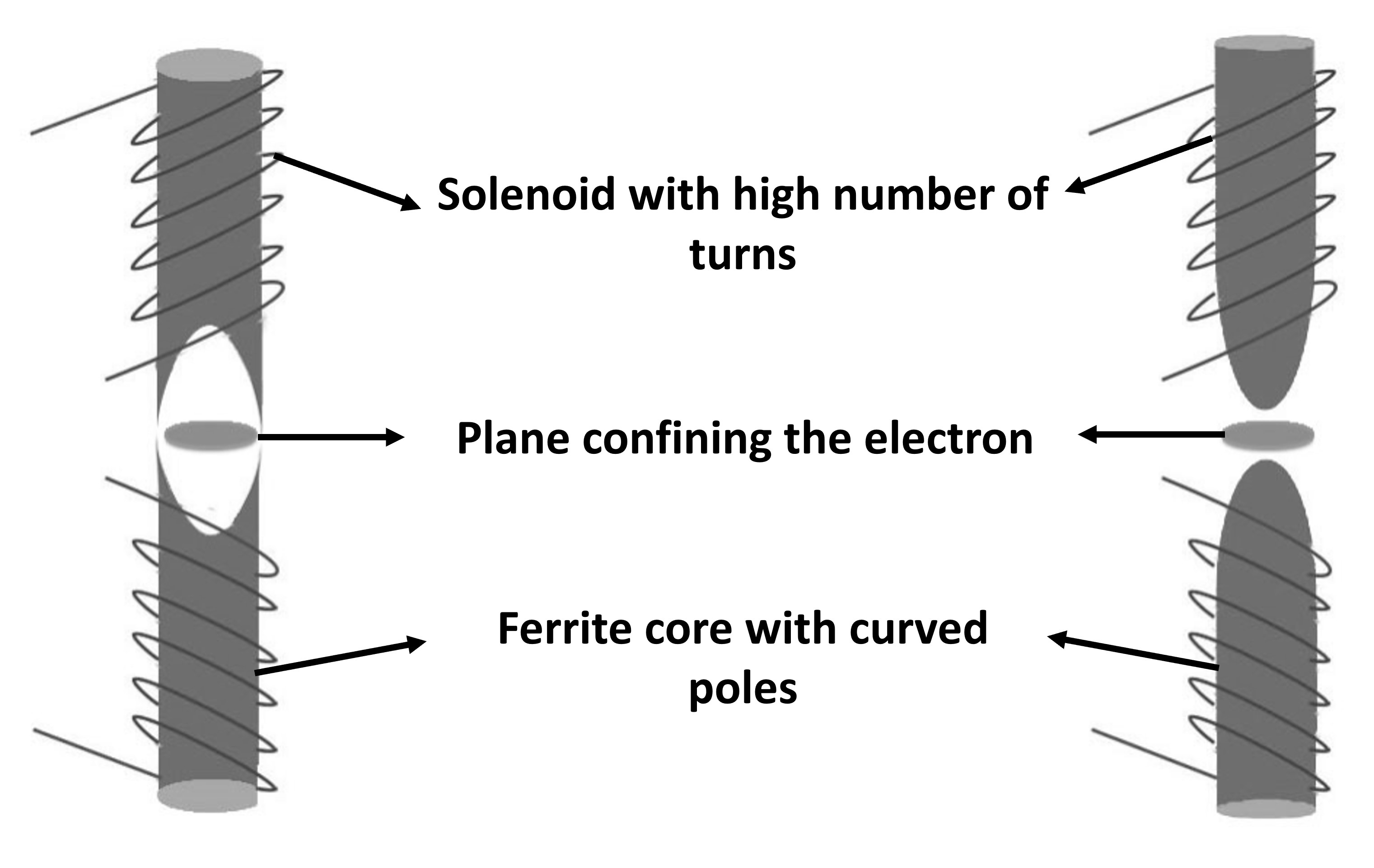}

\caption{The experimental design to attain non-uniform magnetic field. The solenoids with ferrite core having concave (left) and convex (right) pole pieces could be applied to achieve growing ($n>0$) and decaying ($n<0$) magnetic fields respectively.}
\label{fig:experiment}       
\end{figure}

To compare the QSL for electrons in linearly varying magnetic fields against uniform case, we have simulated a laboratory setting. Let us confine the electron at the centre within $1\:\mu m$ in a plane of diameter 1 mm and put it between the solenoids with concave pole pieces. The size of the electron confinement ($1\:\mu m$) is comparable to the scale of quantum dots \cite{Kouwenhoven_2001} or Josephson junction \cite{2019Natur.569...93R}. The solenoid with ferrite core has $100$ turns/cm and carries 1 A current. Assuming a linearly increasing magnetic field profile, the magnetic field will be 0 at the centre and approximately $10^4\:G$ at the edge of the plane with its strength at the periphery of electron's circular plane being $10\:G$.  Mathematically, $|\textbf{B}|=10~G$ at $\rho=0.5\:\mu m=5\times10^5\:pm$ when $n=1$. So, $B_0 = 10/(5\times10^{5})=2\times10^{-5}~G~pm^{-n}$ (using Eq. \ref{eqB}). QSL for spin-up and spin-down electrons are  $~3.2\times 10^{-7}c$ and $~3\times 10^{-7}c$, respectively. For uniform magnetic field, the similar condition with $B_0=10\:G$ provides a QSL of $1.9\times 10^{-7}c$ for both the electrons' spins. Thus, spatially increasing magnetic field leads to higher QSL of electron.

Additionally, spatially varying magnetic fields can be realized through various experimental techniques. For instance, a three-dimensional architecture employing lithographically patterned films with varying thickness and strong perpendicular magnetic anisotropy can generate such fields
\cite{doi:10.1063/1.4749818}. Also,  local magnetic fields with controlled spatial distribution can be induced in materials like graphene or transition metal dichalcogenides by exploiting grain boundaries or topological defects \cite{ACS}. Hence, these methods offer promising avenues for experimentally realizing the increased QSL observed in our model with spatially increasing magnetic fields.

\section{Conclusions}
\label{concl}
We have shown that a non-uniform magnetic field is crucial in attaining higher QSL till $0.6c$. Further, we demonstrate how,
adopting a foundational perspective, we can find the critical magnetic field by using the BB bound. Thus, the BB bound provides a unified framework for examining both relativistic and non-relativistic treatments, enabling a comprehensive understanding of their interrelationship. We further present an experimental blueprint to realize the proposed concepts in a laboratory setting.

Thus, the present endeavor not only accentuates the use of relativistic dynamics to quantify information theory via an increase of QSL in a non-uniform magnetic field, but also probes into the facets of the former using the latter via the use of BB bound to determine the critical magnetic field. Its applications potentially impact a broader spectrum of scientific disciplines, including quantum information and condensed matter physics.
\begin{acknowledgement}
The authors acknowledge ISRA 2023 organising committee for providing a platform to exchange ideas about relativity, astrophysics and quantum physics. 
\end{acknowledgement}
\ethics{Competing Interests}{
The authors have no conflicts of interest to declare that are relevant to the content of this chapter.}


\bibliographystyle{ieeetr}
\bibliography{main_rev2_final}

\begin{thebibliography}{10}

\bibitem{Deffner2017}
S.~{Deffner} and S.~{Campbell}, ``{Quantum speed limits: from Heisenberg{\textquoteright}s uncertainty principle to optimal quantum control},'' {\em Journal of Physics A Mathematical General}, vol.~50, p.~453001, Nov. 2017.

\bibitem{Bremermann}
H.~J. Bremermann, {\em Quantum noise and information Proc. of the 5th Berkeley Symposium on Mathematical Statistics, Probability}, vol.~4.
\newblock Biology, Problems of Health (Berkeley, CA: University of California Press), 1967.

\bibitem{Bekenstein1990}
J.~D. {Bekenstein} and M.~{Schiffer}, ``{Quantum Limitations on the Storage and Transmission of Information},'' {\em International Journal of Modern Physics C}, vol.~1, pp.~355--422, Jan. 1990.

\bibitem{Santini}
P.~Santini, S.~Carretta, F.~Troiani, and G.~Amoretti, ``Molecular nanomagnets as quantum simulators,'' {\em Phys. Rev. Lett.}, vol.~107, p.~230502, Nov 2011.

\bibitem{villamizar}
D.~V. Villamizar and E.~I. Duzzioni, ``Quantum speed limit for a relativistic electron in a uniform magnetic field,'' {\em Phys. Rev. A}, vol.~92, p.~042106, Oct 2015.

\bibitem{SciSris}
S.~Aggarwal, B.~Mukhopadhyay, and G.~Gregori, ``{Relativistic Landau quantization in non-uniform magnetic field and its applications to white dwarfs and quantum information},'' {\em SciPost Phys.}, vol.~11, p.~93, 2021.

\bibitem{1998rqm..book.....S}
P.~{Strange}, {\em {Relativistic Quantum Mechanics}}.
\newblock Cambridge University Press, Cambridge, UK, 1998.

\bibitem{MGSris}
S.~Aggarwal and B.~Mukhopadhyay, {\em Dynamics of relativistic electrons in non-uniform magnetic fields and its applications in quantum computing and astrophysics}, pp.~4362--4373.
\newblock World Scientific, 2023.

\bibitem{mandelstam1945energy}
L.~Mandelstam and I.~Tamm, ``The energy--time uncertainty relation in non-relativistic quantum mechanics,'' {\em J. Phys. USSR}, vol.~9, pp.~249--254, 1945.

\bibitem{MargolousLevitin}
N.~Margolus and L.~B. Levitin, ``The maximum speed of dynamical evolution,'' {\em Physica D: Nonlinear Phenomena}, vol.~120, no.~1, pp.~188--195, 1998.
\newblock Proceedings of the Fourth Workshop on Physics and Consumption.

\bibitem{NJPSris}
S.~Aggarwal, S.~Banerjee, A.~Ghosh, and B.~Mukhopadhyay, ``Non-uniform magnetic field as a booster for quantum speed limit: faster quantum information processing,'' {\em New Journal of Physics}, vol.~24, p.~085001, aug 2022.

\bibitem{deffner}
S.~Deffner and E.~Lutz, ``Quantum speed limit for non-markovian dynamics,'' {\em Phys. Rev. Lett.}, vol.~111, p.~010402, Jul 2013.

\bibitem{Kouwenhoven_2001}
L.~P. Kouwenhoven, D.~G. Austing, and S.~Tarucha, ``Few-electron quantum dots,'' {\em Reports on Progress in Physics}, vol.~64, pp.~701--736, may 2001.

\bibitem{2019Natur.569...93R}
H.~{Ren}, F.~{Pientka}, S.~{Hart}, A.~T. {Pierce}, M.~{Kosowsky}, L.~{Lunczer}, R.~{Schlereth}, B.~{Scharf}, E.~M. {Hankiewicz}, L.~W. {Molenkamp}, B.~I. {Halperin}, and A.~{Yacoby}, ``{Topological superconductivity in a phase-controlled Josephson junction},'' {\em Nature}, vol.~569, pp.~93--98, Apr. 2019.

\bibitem{doi:10.1063/1.4749818}
N.~Rougemaille, A.~T. N'Diaye, J.~Coraux, C.~Vo-Van, O.~Fruchart, and A.~K. Schmid, ``Perpendicular magnetic anisotropy of cobalt films intercalated under graphene,'' {\em Applied Physics Letters}, vol.~101, no.~14, p.~142403, 2012.

\bibitem{ACS}
Z.~Zhang, X.~Zou, V.~H. Crespi, and B.~I. Yakobson, ``Intrinsic magnetism of grain boundaries in two-dimensional metal dichalcogenides,'' {\em ACS Nano}, vol.~7, no.~12, pp.~10475--10481, 2013.
\newblock PMID: 24206002.

\end{thebibliography}
\end{document}